\documentclass[pra,amssymb,twocolumn]{revtex4}

\usepackage{epsfig}
\usepackage{dcolumn}
\usepackage{graphicx}
\usepackage{amsmath, amssymb, times, easybmat}
\usepackage{color}

\newcommand{\be}{\begin{equation}}
\newcommand{\ee}{\end{equation}}
\newcommand{\bea}{\begin{eqnarray}}
\newcommand{\eea}{\end{eqnarray}}

\begin{document}

\title{Acoustic superradiance from an optical superradiance induced vortex in a Bose-Einstein condensate}

\author{Nader Ghazanfari} \email{nghazanfari@msgsu.edu.tr} \affiliation{ Department of Physics, Mimar Sinan Fine Arts University, Bomonti 34380, Istanbul, Turkey}

\author{\"Ozg\"{u}r Esat M\"{u}stecapl{\i}o\u{g}lu}
\affiliation{ Department of Physics, Ko\c{c} University, 34450 Sariyer, Istanbul, Turkey}

\date{\today}

\begin{abstract}
We consider the scattering of an acoustic wave from a vortex induced by an optical superradiance. The vortex is created by pumping a large amount of angular momentum with a Laguerre-Gaussian light beam in an atomic Bose-Einstein condensate. We derive the mean field dynamical equations of the light-superfluid system, and obtain the equations governing the elementary excitation of the system which result in a massless Klein-Gordon equation with source terms. This equation describes the propagation of the sound wave in an effective spacetime. Employing a simplifying draining bathtub model for the vortex, we investigate the scattering of the acoustic wave in the vortex phase and obtain a condition for the acoustic superradiance. We conclude that Laguerre-Gaussian beam induced sudden transition from homogeneous to vortex state in the superfluid leads to a prominent observation of the acoustic superradiance.

\end{abstract}

\maketitle

\section{Introduction}

The achievements in cooling and trapping of the ultracold dilute gases, and developments in controlling their various properties makes it a favourite candidate for simulating different physical systems from solid state to high energy physics \cite{bloch}. A fine control over dilute gases both experimentally and theoretically allows us, by analogy, to analyse the systems which are not easy to study, when dealing with the real one. In this paper, as an example of such efforts we theoretically investigate the possibility of the acoustic superradiance, the analogue version of the Penrose process, which is the extracting of energy from the rotating black holes \cite{penrose,zeldovich}, mutually with the optical superradiance which happens in Bose-Einstein condensates. 

Acoustic superradiance occurs in Bose-Einstein condensation when a sound wave scatters from a vortex with an effective curved spacetime that is the geometry of rotating black holes. Event horizon in such a spacetime exists inside a region called ergoregion, and since the rotating energy of the black hole is located in between event horizon and ergosphere the extraction of energy becomes possible. In other words, in this process, the wave solution of the field equation is scattered from ergoregion with an increase in its amplitude \cite{basak, berti, slatyer, tosi}.

The theoretical framework to study the possible connection between motion of sound waves in a fluid flow and behaviour of a quantum field in a classical gravitational field was constructed by W. G. Unruh in 1981. In his paper \cite{unruh81}, Unruh showed that the equation describing the propagation of the acoustic fluctuation of the velocity potential in a barotropic, inviscid, and irrotational fluid is the same as the equation which governs the propagation of a massless scaler field in a curved spacetime. Since that time the acoustic black holes have gathered a lot of attention. Relatively respectable amount of work  \cite{unruh95, visser, basak, berti, slatyer, tosi, parentani, reznik, fuentes, steinhauer, robertson} have been devoted to make analogies of different features of the black holes among which spontaneous radiation \cite{hawking} and stimulated emissions \cite{penrose, zeldovich, starobinskii, deWitt} are the most engaged properties. 

In the other hand, an ensemble of atoms optically driven above a threshold intensity radiates in the form of superradiance \cite{dick, skribanowitz}. The process occurs in Bose-Einstein condensates \cite{inouye, schneble, fallani, li, deng} for which above a threshold intensity the condensate undergoes another phase transition and rest in vortex state, in case that the incident light carries angular momentum \cite{tasgin}. We aim to observe the acoustic aspect of superradiance along with optical superradiance in presence of an optically driven vortex. For this purpose, we consider a system of bosonic cold atoms cooled down to the condensate state in an elongated trap. The condensate is under a far off-resonant intense beam of laser (in our case Laguerre-Gaussian beams) pumped along the large axis of the trap. The light couples to the atoms and transfers angular momentum to the condensate. The large amount of angular momentum pumped to the condensate to create a vortex throughout a transition from normal state to optical superradiance state provides an opportunity to extract energy from this environment. This process eases the observation of the superradiance in a acoustic superradiance experiment.

Superradiance induced vortex phase has been studied in detail in \cite{tasgin} and we work in this regime searching for the conditions of the acoustic superradiance for the system. Then, the equation of motion for acoustic fluctuation of the velocity field is derived from mean field equations for the condensate order parameter and light modes. The resulting equation is a nonhomogeneous massless scalar field in an effective curved spacetime. The possibility of observing superradiance for the system is discussed throughout the analytical method used in \cite{basak, berti}. This paper is organized as follows: we describe the system writing the Hamiltonian in the first section giving the details of laser and also deriving the equations of motion for condensate and light modes order parameters. In addition we discuss the superradiance induced vortex state qualitatively. In the second section, we introduce the effective geometry of the acoustic black hole, writing the field equation for the phase perturbation of the condensate and discussing the metric and properties of rotating acoustic black hole. In section III, we discuss the possibility of superradiance for the system by calculating the reflection coefficient throughout the scattering of a sound wave from a vortex. Finally we summarise the results in section IV.


\section{Superradiance induced vortex state}
We first review the main results and equations in Ref. \cite{tasgin}, where the generation of a superradiance induced vortex state by angular momentum carrying LG beam is studied. We consider a cigar-shaped Bose-Einstein condensate coupled to a far off-resonant intense laser field along the long axis of the trap. Atoms are interacting via short-range s-wave interaction. The many-body Hamiltonian describing the system is
\bea \nonumber
H&=&\int d^3{\bf r} \hat{\Psi}^{\dagger}({\bf r})H_0\hat{\Psi}({\bf r}) + \sum_m d^3{\bf k} \hbar\omega a^{\dagger}_{km}a_{km} \\ \nonumber &+&  \sum_{m,m'}\int d^3{\bf r} d^3{\bf k} d^3{\bf k'}J\left({\bf k},{\bf k'};{\bf r} \right)\hat{\Psi}^{\dagger}({\bf r})a^{\dagger}_{km}a_{k'm'}\hat{\Psi}({\bf r}) \\  &+&  \frac{1}{2}\int d^3{\bf r} d^3{\bf r'}\hat{\Psi}^{\dagger}({\bf r})\hat{\Psi}^{\dagger}({\bf r'})V({\bf r-r'})\hat{\Psi}({\bf r})\hat{\Psi}({\bf r'}) ,
\eea
where $ H_0$ is the atomic single particle Hamiltonian consisting of a kinetic term and an external trapping potential, $V_{ext}(\bf r)$, $\hat{\Psi}({\bf r})$  and $a_{km}$ are the annihilation operators for atoms and optical field, respectively, and $m$ and $m'$ are labelling the angular momentum for optical modes. Here $V({\bf r-r'})=4\pi\hbar^2a_s/M\delta({\bf r-r'})$ is the two body potential with $a_s$ being the s-wave scattering length and $M$ the mass of a single atom. The effective atom-light coupling coefficients $J\left({\bf k},{\bf k'};{\bf r} \right)$ are given by
\be
J\left({\bf k},{\bf k'};{\bf r} \right) = -\frac{\hbar J^*({\bf k})J({\bf k'})}{\Delta}\Phi^*_{km}({\bf r})\Phi_{k'm'}({\bf r}),
\ee
and determined by the single atom-photon dipole matrix element $g({\bf k})$. Here, $\Delta$ is the detuning frequency, and $\Phi_{km}({\bf r})$ are the mode functions (in our case Laguerre-Gaussian modes) for the light field with the wave number ${\bf k}$. These mode functions are given as 
\bea \label{LGmodes}\nonumber
\Phi_{km}({\bf r})&=&\phi_m(r)e^{im\phi}e^{ikz},\\ 
&=& \frac{1}{\sqrt{\pi}}\left(\frac{r}{a_m}\right)^m e^{-r^{2}/2a_m}e^{im\phi}e^{ikz}.
\eea

The laser beam has a width of $a_m$ and carries $m\hbar$ units of orbital angular momentum. We write the Heisenberg equation of motion for four annihilation operators, and apply mean field approximation, whereby the field operators are replaced by $c$-numbers. In particular, we replace $\hat{\Psi} \rightarrow \psi$, $a_{-k_01}\rightarrow \alpha_1$, $a_{-k_00}\rightarrow \alpha_2$, $a_{k_00}\rightarrow \alpha_3$, and $a_{k_01}\rightarrow \alpha_L$ that lead us to expressions 

\bea \label{heisenberg1}
i\partial_t \psi &=& \left[-\frac{\hbar}{2M}\nabla^2 + V + \frac{4\pi a_s \hbar}{M} |\psi|^2 - J_l\right]\psi \\ \nonumber
i \partial_t\alpha_1 &=& \left[-\Delta_1 - 2U_0 I^{(11)}_{--}\right]\alpha_1 - U_0 I^{(11)}_{-+}\alpha_L \\ \label{heisenberg2} i \partial_t\alpha_2 &=& \left[ - \Delta_2 - 2U_0 I^{(00)}_{--}\right]\alpha_2 - U_0 I^{(01)}_{-+}\alpha_L \\\nonumber  i \partial_t\alpha_3 &=& \left[- \Delta_3 - 2U_0 I^{(00)}_{++}\right]\alpha_3 - U_0 I^{(01)}_{++}\alpha_L,
\eea
where $U_0 = J_k^2/\Delta$, $\Delta_i$'s are the end-fire mode frequencies in the rotating frame at frequency $\omega_0$, $J_l$ is the light-atom coupling
\bea \nonumber
J_l &=&  2U_0 \big\{ |\alpha_L |^2 |\Phi_{k_01} |^2 + |\alpha_1 |^2 |\Phi_{-k_01} |^2 \\ \nonumber &+& |\alpha_2 |^2 |\Phi_{-k_00} |^2 + |\alpha_3 |^2 |\Phi_{k_00} |^2 \big\}  \\ \nonumber &+& U_0 \big\{\alpha_L  \alpha^*_1\Phi^*_{-k_01}\Phi_{k_01} + \alpha^*_2\Phi^*_{-k_00}\Phi_{k_01} \\  &+& \alpha^*_3\Phi^*_{k_00}\Phi_{k_01} + c.c.\big\},
\eea and
\be
I^{mm'}_{\sigma\gamma}= \int d^3{\bf r} \Phi^*_{\sigma k_0m}({\bf r})\Phi_{\gamma k_0m'}({\bf r})|\psi({\bf r})|^2.
\ee
Here $\sigma$, and $\gamma = \pm1$ label the sign of the wavevectors with amplitude $k_0$. 

The Eq. (\ref{heisenberg1}) is the Gross-Pitaevskii equation for a condensate coupled with a laser beam. The Eqs. (\ref{heisenberg1}) and (\ref{heisenberg2}) have been solved numerically in \cite{tasgin} and the optical superradiance has been observed. In that paper, Tasgin et al. illustrate the dynamics of the transition from a condensate at its non-rotating ground state to a normal superradiance and then a rotatory superradiance and finally a superradiance induced vortex phase for the condensate. We will work in this phase where after a certain density of laser beam two transitions happens and the superradiance with a topological vortex coexist. According to the dynamics of the transition discussed in Ref. \cite{tasgin}, the mean photon number in mode $\alpha_1$ decreases dramatically, $\alpha_2$ remains unchanged but very small and only the $\alpha_3$ mode survive in this phase where it increases sharply when superradiance take place. Assuming that the system reside in this regime we aim to find the possibility of observing the acoustic superradiance along with the optical superradiance when a sound wave scatters from the vortex.


\section{Acoustic black hole: effective geometry}

In order to investigate the possibility of observing the acoustic superradiance we need to write the equation governing the propagation of the acoustic fluctuation of the velocity potential in the effective geometry created by the vortex. We start from Gross-Pitaevskii Eq. (\ref{heisenberg1}) and Express the condensate order parameter in terms of its amplitude and phase, i.e $\psi({\bf r},t)=\sqrt{\rho({\bf r},t)} e^{iS({\bf r},t)}$, where $\rho=|\Psi|^2$. This leads us us to two equations for real and imaginary parts of the Eq. (\ref{heisenberg1}) as
\bea
\label{densityequation} \partial_t \rho({\bf r},t)&=&-\frac{\hbar}{M}\left[ \nabla\rho\cdot \nabla S + \rho \nabla^2 S \right] \\ \nonumber  \partial_t S({\bf r},t) &=& \frac{\hbar}{2M}\frac{1}{\sqrt{\rho}}\nabla^2\sqrt{\rho} - \frac{\hbar}{2M}|\nabla S|^2 \\ \label{phaseequation} &-& V_{ext}-\frac{4\pi a_s\hbar}{M}\rho+J_{l}
\eea
By linearizing the Eqs. (\ref{densityequation})-(\ref{phaseequation}) and (\ref{heisenberg2}) for density, phase, and the light-atom coupling around the background values $\rho_0$, $S_0$, and $J_0$ in the optical superradiance induced vortex phase, as
\bea
\label{linearization} \rho=\rho_0+\rho_1, \hspace*{.2cm} S=S_0+S_1, \hspace*{.5cm} J_l&=&J_0+J_1,
\eea
where $J_0$, and $J_1$ are 
\bea \nonumber \label{g0} J_0 &=&  2U_0 \left[ |\alpha_L |^2 |\Phi_{k_01} |^2 + |\alpha_3 |^2 |\Phi_{k_00} |^2\right] \\ \nonumber &+& U_0 \alpha_L \left[\alpha_3\Phi_{k_00}\Phi^*_{k_01} + c.c.\right], \\ \nonumber \label{g1}  
J_1 &=& U_0\alpha_L \big[ \delta\alpha_1 \Phi_{-k_01}\Phi^*_{k_01} + \delta\alpha_2 \Phi_{-k_00}\Phi^*_{k_01} \\ \nonumber &+& \delta\alpha_3 \Phi_{k_00}\Phi^*_{k_01}+ c.c. \big],
\eea 
we obtain
\bea
\label{linearizeddensity} \partial_t \rho_1&=&-\frac{\hbar}{M}\left[\nabla\cdot( \rho_0\nabla S_1 )- \nabla\cdot(\rho_1\nabla S_0)\right], \\ \label{linearizedphase} \partial_tS_1 &=& -\frac{\hbar}{M}\nabla S_0\cdot \nabla S_1 - \frac{4\pi a\hbar}{M}\rho_1 + J_1.
\eea
It should be noted that in Eq. (\ref{linearizedphase}) we have neglected the quantum pressure term, 
\bea \nonumber \frac{\hbar^2}{2M}\left( \frac{1}{2\sqrt{\rho_0}}\nabla\frac{\rho_1}{\sqrt{\rho_0}}-\frac{\rho_1}{2\rho_0^{3/2}}\nabla^2\sqrt{\rho_0} \right). \eea
The Eqs. (\ref{linearizeddensity}) and (\ref{linearizedphase}) should be solved together with the linearized equations of the modes $\alpha_1$, $\alpha_2$, and $\alpha_3$.
\bea \label{linerarizedalpha1} 
i\partial_t \delta\alpha_1 &=& - \Delta_1 \delta\alpha_1 - 2 \frac{U_0}{\hbar} I^{11}_{--} \delta\alpha_1,\\ \label{linerarizedalpha2} 
i\partial_t \delta\alpha_2 &=& - \Delta_2 \delta\alpha_2 - 2 \frac{U_0}{\hbar} I^{00}_{--} \delta\alpha_2, \\ \label{linerarizedalpha3} 
i\partial_t \delta\alpha_3 &=& - \Delta_3 \delta\alpha_3 - 2 \frac{U_0}{\hbar} I^{00}_{++} \delta\alpha_3.
\eea

In order to write the Eqs. (\ref{linearizeddensity}) and (\ref{linearizedphase}) in a compact form we can use the definitions for the the background flow velocity ${\bf v}$, and the speed of sound $c$ in a condensate,
\bea \label{velocity-sound}
{\bf v}=\frac{\hbar}{M}\nabla S_0, \hspace*{.5cm}
c=\frac{\hbar}{M}\sqrt{4\pi a_s\rho_0}.
\eea
We assume that the background density $\rho_0$ is constant, thus the speed of sound. Now Eqs. (\ref{linearizeddensity}) and (\ref{linearizedphase}) can be combined and rewritten in a single equation for sound waves as
\bea \label{klein-gordon}
\frac{1}{\sqrt{-g}}\partial_\mu\left( \sqrt{-g} g^{\mu\nu}\partial_\nu S_1\right) = - \partial_tJ_1 - \nabla\cdot({\bf v}J_1),
\eea
where $\mu$ and $\nu = 0,1,2$, and $g^{\mu\nu}$, the inverse metric tensor is obtained as
\begin{align}
g^{\mu\nu} = \frac{1}{c^2} \left(
\begin{array}{ccccc}
-1 & \vdots &-v_r & -\frac{v_\theta}{r}  \\
\cdots & \cdot &\cdots\cdots & \cdots\cdots  \\
-v_r & \vdots & c^2-v_r^2 & -\frac{v_rv_\theta}{r} \\
-\frac{v_\theta}{r} & \vdots & -\frac{v_rv_\theta}{r} & \frac{c^2- v_\theta^2}{r^2} \\
\end{array} \right). \hspace*{1cm} 
\end{align}
The Eq. (\ref{klein-gordon}) is a non-homogeneous massless Klein-Gordon equation in curved spacetime, for which $g=det(g_{\mu\nu})$ and the metric tensor in polar coordinates is defined as
\begin{align}
g_{\mu\nu} = \left(
\begin{array}{ccccc}
-(c^2-v^2) & \vdots & -v_r & -rv_\theta  \\
\cdots\cdots & \cdot &\cdots & \cdots  \\
-v_r & \vdots & 1 & 0 \\
-rv_\theta & \vdots & 0 & r^2 \\
\end{array} \right).
\end{align}

This metric governs the propagation of the fluctuations (sound waves) and depends on the velocity field, and speed of sound thus the density of the condensate. Even though the dynamics of the atomic Bose-Einstein condensates is driven from a non-relativistic equation, the behaviour of the sound waves is specified by a relativistic equation in a curved space time \cite{unruh81, visser}. The homogeneous form of the Eq. (\ref{klein-gordon}) introduced by Unruh \cite{unruh81} for a barotropic, inviscid, and irrotational fluid establishes the connection between the propagation of the scalar field in classical gravitational field and the wave sounds in curved spacetime. We will discus later, but it is worth to note that the optical superradiance does not affect the effective curved spacetime.  

To observe the certain properties of the spacetime it is better to write the metric from the metric tensor
\bea 
ds^2 &=& g_{\mu\nu}dx^\mu dx^\nu \\ \nonumber &=& (v^2-c^2)dt^2-2v_rdrdt-2rv_\theta d\theta dt + dr^2+r^2d\theta^2.
\eea
The ergoshpere radius can be easily found from this metric and it is exactly where the temporal component of metric, i.e.  $g_{00}$ changes sign. However to find the event horizon one need to apply a coordinate transformation of form
\bea \nonumber
dt &\longrightarrow &  dt - \frac{v_r}{c^2-v_r^2}dr, \\
d\theta &\longrightarrow & d\theta - \frac{v_rv_\theta}{r(c^2-v_r^2)}dr, 
\eea
which results in the metric
\bea \nonumber
ds^2  &=& -(c^2-v^2)dt^2 \\ &+& \left( \frac{c^2}{c^2-v_r^2}\right) dr^2 - 2rv_\theta d\theta dt+ r^2d\theta^2.
\eea
The metric in the new coordinates has an obvious singularity at radial component which gives the radius of event horizon. 

Now we need to specify the form of the flow velocity. The spatial profile of the superradiance generated vortex is numerically determined in Ref. \cite{tasgin}. For our analytical examination, we simply choose a draining bathtub profile, which is typical description of rotating acoustic black holes. This model was first used in \cite{visser} for rotating acoustic black holes which is a $(2+1)$-dimensional flow with a sink or source at the origin. We assume that the density and velocity have pure radial dependency. The continuity equation with irrotationality and incompressibility of the flow lead us to write the velocity field as
\bea \label{velocityfield}
{\bf v} = \frac{A}{r}\hat{e}_r + \frac{B}{r}\hat{e}_{\theta},
\eea
where $A$, and $B$ are constants and can be defined in terms of the black hole properties. The field equations derived from conservation laws mentioned above also result in a position independent background density $\rho_0$ throughout the flow which automatically gives the constant speed of sound according to Eq. (\ref{velocity-sound}). Having the velocity field defined by Eq. (\ref{velocityfield}), it can be easily checked that the ergosphere and event horizon are formed at $r_{erg}$, and $r_h$, respectively
\bea
r_{erg}=\frac{\sqrt{A^2+B^2}}{c}, \hspace*{.5cm}
r_h=\frac{|A|}{c}.
\eea
The sign of $A$ is of no importance in determining the ergoregion, but it make difference when dealing with event horizon. For positive $A$ the past event horizon is defined that means we work with an acoustic white hole , while for negative $A$ the future event horizon is defined this time that means we work with an acoustic black hole. We choose $A=-ac$, and $B=a^2\Omega$ where $a$ is the radius of the event horizon, and $\Omega$ is the angular velocity of the rotating black hole \citep{tosi}. We will see that the growth of the ergosphere with increasing angular velocity of black hole will increase the amount of acoustic superradiance from vortex. Now, we write the Klein-Gordon equation introduced above on this background in more explicit form of
\bea \label{explicitKG} \nonumber
&&\bigg[-\frac{1}{c^2} \partial^2_t + \frac{2a}{r}\partial_t\partial_r - \frac{2\Omega a^2}{c^2r^2}\partial_t\partial_{\theta}+ \left( \frac{c^2r^2-\Omega^2a^4}{c^2r^4} \right)\partial^2_{\theta} \\ \nonumber &&+ \left( 1-\frac{a^2}{r^2} \right)\partial_r^2 + \frac{2a^3\Omega}{cr^3}\partial_r\partial_{\theta} + \frac{a^2+r^2}{r^3}\partial_r \\ &&- \frac{2a^3\Omega}{cr^4}\partial_{\theta} \bigg]S_1({\bf r},t) = - \partial_t J_1({\bf r},t)-\nabla\cdot\left({\bf v}J_1({\bf r},t)\right).
\eea
The homogeneous version of the equation above has been solved analytically \citep{basak, berti, slatyer}, and numerically \citep{tosi} and the superradiance has been observed. Since the superradiance is the extraction of energy from vortex, the problem can be reduced to find the reflection and transmission coefficients and discuss the possibility of finding a reflection probability greater than unity. The analytical method with some transformations gives the result relatively easily, but the numerical solution is not as easy. The method developed in \citep{teukolsky} reduces the the Klein-Gordon equation to a set of first-order equations by defining two conjugate fields; however the resulting set of equations itself requires many numerical calculations. The method has been implied in \cite{tosi, tosi2} and the superradiance state has been discussed in details. In the case of our system the problem becomes even more difficult since the equation must be solved along with the linearized equations of motions for $\delta\alpha_1$, $\delta\alpha_2$, and $\delta\alpha_3$. However, considering the dynamics of the condensate throughout the optical superradiance, since the $\alpha_1$ and $\alpha_2$ modes nearly vanish in this phase, one can neglect the contribution from these modes. Therefore, the analytical calculations reduce to solving the Eq. (\ref{klein-gordon}), where the source term is determined by Eq. (\ref{linerarizedalpha3}).


\section{Superradiance}
The scattering properties of a sound wave from a superradiance induced vortex is described by analysing the massless Klein-Gordon equation (\ref{explicitKG}). We separate the phase fluctuations $S_1$ into its variables by substitution of
\bea 
S_1(t;r,\theta)= R(r)S(t;\theta) = R(r)e^{i(n\theta-\omega t)}, 
\eea
which results in a nonhomogeneous second order differential equation for the perturbed phase. Here n is the azimuthal quantum number with respect
to the axis of rotation, and $\omega$ is the sound wave frequency. We divide both sides of the resulting equation by factor $l=1-a^2/r^2$ to obtain more familiar form of
\bea \label{radialequation}
\frac{d^2R(r)}{dr^2} + P(r)\frac{dR(r)}{dr} + Q(r)R(r) = G(t;r,\theta),
\eea
where
\bea \label{PQdef} \nonumber
P(r) &=& \frac{1}{cr(r^2-a^2)}\big[ c(a^2+r^2) + 2i\omega a r^2 - 2 i n\Omega a^3  \big], \\ 
Q(r) &=& \frac{1}{c^2r^2(r^2-a^2)}\big[ n^2\Omega^2a^4 +\omega^2 r^4 \\ \nonumber &-& n^2c^2r^2 - 2n\omega \Omega a^2r^2-2inc\Omega a^3 \big].
\eea
The source term in Eq. (\ref{explicitKG}) includes the time and spatial derivatives, where the time dependency of $J_1$ in optical superradiance state is governed by Eq. (\ref{linerarizedalpha3}). We can write $J_1({\bf r},t)= \tilde{G}(t;\theta)J(r)$, where $\tilde{G}(t;\theta)$ has the simple time dependency of the form $e^{i\omega_l t}$, with $\omega_l= \Delta_3 + 2U_0 I^{00}_{++}/\hbar$. Now, the source term in Eq.~(\ref{radialequation}) can be then conveniently expressed as 
\bea \nonumber
G(t;r,\theta) &=& \frac{-1}{lS(t;\theta)}\bigg[ \partial_t \tilde{G}(t;\theta) + \frac{\Omega a^2}{r^2}\partial_\theta \tilde{G}(t;\theta) \\ &-&  \tilde{G}(t;\theta) \frac{ca}{r} \frac{d}{dr} \bigg]J(r).
\eea
At the end of the previous section we discussed the dynamics of the condensate throughout the optical superradiance and emphasized that $J_1$ is a very simple expression only carrying $\phi_{0}$, and $\phi_{1}$ modes of Laguerre-Gaussian beam since only $\alpha_3$ survives in this phase. Thus, the position derivatives of $J$ gives a simple expression,
\bea \nonumber
\frac{dJ(r)}{dr} &=& \frac{d}{dr}\left[\phi_{0}\phi_{1}\right] = \frac{d}{dr}\left[\left(\frac{r}{a_1}\right)e^{\frac{-r^2}{2a_{red}^2}}\right] \\ &=& \left( \frac{1}{r}-\frac{r}{a_{red}^2} \right)J(r),
\eea
where $a_{red}^2=a_0^2a_1^2/(a_0^2+a_1^2)$ is the reduced width of the Laguerre-Gaussian beam.

Now we introduce a new coordinate $\tilde{r}$, known as tortoise coordinate \cite{wald} and use the definition $dr=ld\tilde{r}$ which lead us to a transformation relation of
\bea \label{transformation}
\tilde{r} = r - \frac{a}{2}\ln \left| \frac{r+a}{r-c}\right|.
\eea
Note that this transformation maps the horizon at $r_h=a$ to $\tilde{r} \longrightarrow -\infty$, and also maps $r \longrightarrow \infty$ to $\tilde{r} \longrightarrow \infty$. These mapping will be important when we check the behaviour of the system at its asymptotic points. In order to investigate the possibility of the acoustic superradiance for our nonhomogeneous Klein-Gordon equation we follow a formal way used for homogeneous one in \cite{basak,berti,slatyer}, in which the superradiance is determined by the reflection and transmission coefficients. In order to facilitate the calculations of these coefficients we write the second order differential equation (\ref{radialequation}) in the form of the usual Schrodinger equation. We set $R(r)=K(r)F(r)$, which along with the coordinate transformation give us,
\bea \label{secorddiffeq}
\frac{d^2F(\tilde{r})}{d\tilde{r}^2}+ D(r)\frac{dF(\tilde{r})}{d\tilde{r}} + W(r)F(\tilde{r})  = \frac{l^2}{K(r)}G(t;r,\theta),
\eea
where
\bea \label{Weq}
W(r) = \frac{l^2}{K(r)}\left[ \frac{d^2K(r)}{dr^2} + P(r)\frac{dK(r)}{dr} + Q(r)K(r)\right].
\eea
Here $K(r)$ is obtained from the elimination of the first derivative term from differential equation Eq. (\ref{secorddiffeq}), i.e., by equating $D(r)$ to zero,
\bea 
\frac{dK(r)}{dr} + \frac{1}{2}\left[P(r)+ l\frac{d}{dr}\left(\frac{1}{l}\right)\right]K(r)=0.
\eea
By solving this equation for $K(r)$ and substituting $P(r)$ from definitions Eq. (\ref{PQdef}) one can obtain
\bea \nonumber
K(r)&=&\sqrt{r}exp\bigg\{\left(\frac{ina\Omega}{c}+1\right)\ln\left(\frac{1}{r}\right) \\  &-& \frac{ia(\omega-n\Omega)}{2c}\ln\left[ c^2(r^2-a^2)\right]\bigg\}.
\eea 
We substitute $K(r)$ in Eq. (\ref{Weq}) to obtain the 
\bea
 \nonumber W(r) &=& \frac{1}{c^2}\left( \omega -\frac{n\Omega a^2}{r^2} \right)^2 - \frac{1}{r^2}\left(n^2 -\frac{1}{4} \right) \\ &+& \frac{a^2}{r^4}\left(n^2- \frac{3}{2} \right) + \frac{5a^4}{4r^6}.
\eea
Eventually, the Eq. (\ref{secorddiffeq}) becomes
\bea \label{modifieddiffeq}
\frac{d^2F(\tilde{r})}{d\tilde{r}^2} + W(r)F(\tilde{r})  = \frac{l^2}{K(r)}G(t;r,\theta),
\eea

We scale the radial coordinate with length of the horizon, i.e. $r_{new}=r/a$, and the frequencies with sound wave frequency, $\omega_{new}=a\omega/c$, and $\Omega_{new}=a\Omega/c$. However to avoid using the new index we drop it and continue with writing as old parameters. 

\begin{figure}
\begin{center}
\includegraphics[width=8cm]{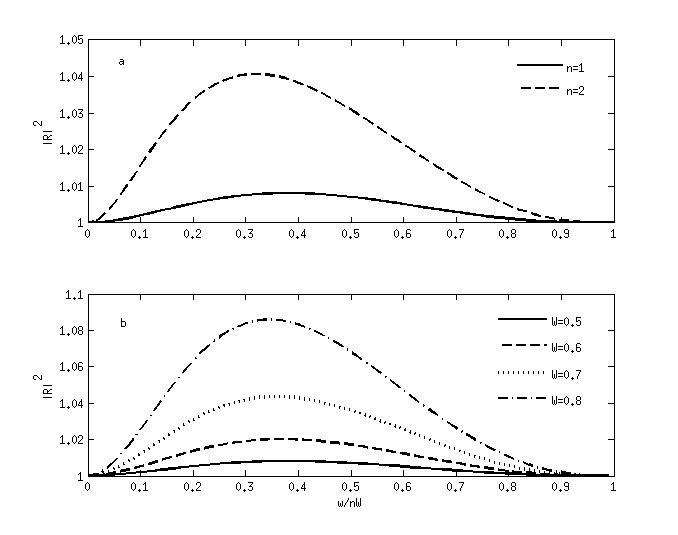}
\end{center}
\caption{The reflection amplitude $|R|^2$ as a function of $\omega/n\Omega$. (a) The reflection coefficient for $n=1,2$. The amplification in reflection coefficient increases by increasing the orbital angular momentum. (b) The reflection coefficient for different values of analogue black hole angular frequency $\Omega$. The increase in reflection amplitude for large values of the angular frequency is more prominent than that for the large values of the orbital angular momentum.}
\label{reflectionfig}
\end{figure}

In the asymptotic region when $r$, and $\tilde{r} \longrightarrow +\infty$ the terms with $1/O(r)$ in $W(r)$ vanishes, and only the term with $\omega$ survives. The source term also vanished in this region due to the Gaussian term in $J(r)$. Thus, the Eq. (\ref{modifieddiffeq}) becomes
\bea
\frac{d^2F(\tilde{r})}{d\tilde{r}}+ \omega^2F(\tilde{r}) = 0,
\eea
which can be readily solved and written as a combination of incident wave and reflected one
\bea
F(\tilde{r}) = R e^{ i\omega \tilde{r} } + e^{- i\omega \tilde{r}},
\eea
so $R$ is the reflection coefficient. Now let us check the behaviour of the differential equation around horizon when $r\longrightarrow 1$, and $\tilde{r}\longrightarrow -\infty$. In this region the non-homogeneous term vanishes due to $l$ which is zero at horizon. Thus, the Eq. (\ref{modifieddiffeq}) reduces to
\bea
\frac{d^2F(\tilde{r})}{d\tilde{r}}+ \left( \omega- n\Omega \right)^2F(\tilde{r}) = 0,
\eea
for which the solution can be written in terms of transmission wave as
\bea
F(\tilde{r})=T e^{i(\omega-n\Omega)\tilde{r}}.
\eea
where $T$ is the transmission coefficient. From conservation law for current density we obtain the relation between reflection and transmission coefficients
\bea \label{superradiance}
\vert R\vert^2 = 1+ \left( \frac{n\Omega}{\omega} -1 \right)\vert T \vert^2,
\eea
which leads us to the famous relation first obtained by Zeldovich \cite{zeldovich} for the scattering of a electromagnetic wave with an orbital momentum n and frequency $\omega$ from a cylinder rotating with an angular frequency $\Omega$. This relation indicates that for $\omega < n\Omega$, an amplifications occurs in reflection coefficient throughout the scattering which is an evidence for possibility of the acoustic superradiance in our analogue system. Here $\Omega$ is related to the amount of angular momentum pumped to the condensate to create a superradiance induced vortex state. Although the Eq. (\ref{superradiance}) determines the region for superradiance it does not give the details of the scattering and its dependency on the sound wave frequency. Therefore we need a more detailed investigation of the reflection and transmission coefficients by solving the differential equation (\ref{radialequation}) explicitly. previously, in this section in order to get a condition for superradiance, we used the transformation $R(r)=K(r)F(r)$ on Eq. (\ref{radialequation}) and investigated the resulting equation in asymptotic limit where the source term has no effect. Similarly now we apply another transformation of the form $R(r)=r^{3/2}K(r)X(r)$ and applied the result of the asymptotic limit to find the reflection and transmission coefficients explicitly. Thus, the Eq. (\ref{radialequation}) reduces to a homogeneous differential equation
\bea \label{RiemannEq1}  
&& x(x+1)\frac{d^2X(x)}{dx^2} + (2x+1) \frac{dX(x)}{dx} \\ \nonumber &+& \frac{1}{4}\left( \frac{u_1^2}{x} + \frac{1}{x+1} + 1 - u_2^2 + \frac{a^2\omega^2}{c^2}x\right)X(x)=0,
\eea  
where
\bea \nonumber
u_1 &=& \frac{a\omega}{c}\left(\frac{n\Omega}{\omega}-1\right) ,\\
u_2^2 &=& n^2 + \frac{2a^2\omega^2}{c^2}\left( \frac{n\Omega}{\omega} -1 \right).
\eea   
The Eq. (\ref{RiemannEq1}) is similar to the equation used by Starobinskii \cite{starobinskii} to calculate the details of the amplification which occurs for reflection coefficient during a superradiance from a rotating black hole. For sound waves with a wavelength $\lambda$ much larger than the radius of the horizon $a$, the Eq. (\ref{RiemannEq1}) reduces to the Riemann-Papparitz equation \cite{abramowitz} with two regular singular point at $x=0,-1$, which has been investigated in details in \cite{starobinskii} for rotating black holes and used in \cite{basak1} for the acoustic superradiance from a vortex. Therefore, we assume that $a<<\lambda$ which result in  
\bea \label{RiemannEq2}  \nonumber
&& x(x+1)\frac{d^2X(x)}{dx^2} + (2x+1) \frac{dX(x)}{dx} \\  &+& \frac{1}{4}\left( \frac{u_1^2}{x} + \frac{1}{x+1} + 1 - u_2^2 \right)X(x)=0.
\eea
The reflection coefficient can be calculated from this equation through the transformation of this differential equation to a hypergeometric form with known solutions \cite{starobinskii,basak1}. The solutions near the horizon are the superposition of ingoing and outgoing parts, from which one can obtain the reflection and transmission coefficients
\bea \label{reflection}
|R|^2 = 1 + \frac{2a\omega}{cu_2|y_1-iy_2|^2}\left(\frac{n\Omega}{\omega}-1\right),
\eea  
where
\bea
y_1 &=& \frac{\Gamma(1-iu_1)\Gamma(u_2)}{\Gamma(\frac{u_2}{2}-i\frac{u_1}{2})\Gamma(1 + \frac{u_2}{2}-i\frac{u_1}{2})}, \\
y_2 &=& \frac{\Gamma(1-iu_1)\Gamma(-u_2)}{\Gamma(\frac{-u_2}{2}-i\frac{u_1}{2})\Gamma(1-\frac{u_2}{2}-i\frac{u_1}{2})}.
\eea
The transmission coefficient can be obtained easily from comparing the Eq. (\ref{reflection}) with the Eq. (\ref{superradiance}). A detailed analysis of the reflection coefficient in Eq. (\ref{reflection}) reveals the advantage of the investigation of the acoustic superradiance from an optical superradiance induced vortex. We demonstrate that in figure \ref{reflectionfig}, which illustrates the amplification of reflection coefficient in Eq. (\ref{reflection}) thorough the scattering of a sound wave from a vortex. Figure \ref{reflectionfig}(a) compares the magnitude of the acoustic superradiance for two different modes, i.e. $n=1,2$, while figure \ref{reflectionfig}(b) shows the magnitude of the superradiance for the first orbital angular momentum $n=1$ with different angular frequencies, i.e., $\Omega$ values. Comparing the figures \ref{reflectionfig}(a) and (b) exhibits that the superradiance for large values of angular frequency is more evident than the superradiance for the large values in orbital angular momentum. Since in our system the large amount of angular momentum is pumped to the condensate to obtain a optical superradiance induced vortex the observation of acoustic superradiance throughout an experiment would be more prominent.


\section{Summary and Discussion}
For a superradiance phase with an induced topological vortex in an atomic Bose-Einstein condensate we theoretically reveal the acoustic superradiance. This phenomenon is the analogue of the Penrose process for rotating black holes \cite{penrose, zeldovich}. The vortex state and superradiance phase are created by a sudden transfer of an incident angular momentum to the condensate \cite{tasgin}. In order to observe the optical superradiance mutually with the acoustic superradiance we assume that the condensate has gone through a phase transition to the optical superradiance induced vortex state. Since the optical superradiance phase happens with pumping a large amount of angular momentum around vortex core, the extracting of the energy from ergoregion becomes easier. The effect of phase transition does not appear in the effective geometry of the vortex, but appears as a nonhomogeneous part in the Klein-Gordon equation describing the propagation of the sound wave in the introduced effective geometry, which is the geometry of a rotating black hole. The draining bathtub model fits the velocity field created by the optical superradiance. This model introduces an event horizon and an ergoregion. It is shown that the existence of the event horizon is not necessary to observe the Penrose process \cite{slatyer}. However, since the optical superradiance happens inside the event horizon the use of a fitting velocity field becomes essential. 

The acoustic superradiance is determined for the vortex state as the amplification of the reflection coefficient which becomes larger than unity \cite{basak, berti, slatyer} in this phase. We analytically show that the optical superradiance happens inside the effective event horizon and it does not affect the acoustic superradiance. The solutions of the nonhomogeneous Klein-Gordon equation in the asymptotic region gives the conservation law for the current density, which lead us to the acoustic superradiance mutually with optical superradiance. We obtain the same condition already introduced in \cite{zeldovich,basak,berti}. We also calculate the reflection coefficient and show that the acoustic superradiance becomes more prominent for our system since there is a large amount of angular momentum in vortex state induced by optical superradiance. The acoustic superradiance condition exhibits that it happens for non-zero modes when the vortex angular frequency becomes larger than the sound wave propagation frequency. The full numerical solution of this problem would be illuminating to reveal the details of the superradiance transitions more explicitly.

\acknowledgements
N.G. thanks to T\"{U}B\.{I}TAK for the support.

\end{document}